# Anomalous Acoustic Plasmon Mode from Topologically Protected States

Xun Jia[1,2], Shuyuan Zhang[1,2], Raman Sankar[3,4], Fang-Cheng Chou[4], Weihua Wang[1], K. Kempa[5], E. W. Plummer[6], Jiandi Zhang[6], Xuetao Zhu[1,2*], Jiandong Guo[1,2,7*]

1. Beijing National Laboratory for Condensed Matter Physics and Institute of Physics, Chinese Academy of Sciences, Beijing 100190, China.

2. School of Physical Sciences, University of Chinese Academy of Sciences, Beijing 100049, China.

3. Institute of Physics, Academia Sinica, Taipei, 11529, Taiwan.

4. Centre for Condensed Matter Sciences, National Taiwan University, Taipei, 10617, Taiwan.

5. Department of physics, Boston College, Chestnut Hill, MA 02467, USA.

6. Department of Physics and Astronomy, Louisiana State University, Baton Rouge, Louisiana 70808, USA.

7. Collaborative Innovation Center of Quantum Matter, Beijing 100871, China.

*To whom correspondence should be addressed.

 E-mail: xtzhu@ iphy.ac.cn (X.Z.); jdguo@iphy.ac.cn (J.G.)

**Abstract**:

**Plasmons, the collective excitations of electrons in the bulk or at the surface, play an important role in the properties of materials, and have generated the field of "plasmonics". We report the observation of a highly unusual acoustic plasmon mode on the surface of a three-dimensional topological insulator (TI), $Bi_2Se_3$, using momentum resolved inelastic electron scattering. In sharp contrast to ordinary plasmon modes, this mode exhibits almost linear dispersion into the second Brillouin zone and remains prominent with remarkably weak damping not seen in any other systems. This behavior must be associated with the inherent robustness of the electrons in the TI surface state, so that not only the surface Dirac states but also their collective excitations are topologically protected. On the other hand, this mode has much smaller energy dispersion than expected from a continuous media excitation picture, which can be attributed to the strong coupling with surface phonons.**

Plasmon is a quantum of collective oscillations of charge density due to the restoring force arising from the long-range Coulomb interactions in a solid, predicted in 1951 by Pines and Bohm [1]. The concept was extended to surfaces and interfaces in the form of surface plasmons by Ritchie in 1957 [2]. The properties of plasmons have been studied for over 60 years, and has spawned the field of plasmonics and many technological applications. A seminal feature of any plasmon mode, which is important for applications, is its lifetime. In ordinary materials, plasmons usually appear in only small momentum ranges, because they can interact with the single-particle excitations (so-called Landau damping) [3,4] when the plasmon energy and momentum overlap the electron-hole pair continuum (EHPC). Even for plasmon with energy and momentum outside the EHPC there are other damping channels [5,6], such as impurity scattering that increases with increasing momentum. This is true for all plasmon modes, originating from either electron gases in a parabolic band or Dirac massless electrons in a linear band. One example is the two-dimensional (2D) plasmon mode from Dirac electrons in graphene [7], which shows strong damping and diminishes before merging into the EHPC at a small momentum compared to the Brillouin zone (BZ) boundary. In contrast, we demonstrate that an anomalous acoustic surface plasmon mode from the Dirac electrons on the surface of a prototypical three-dimensional (3D) topological insulator (TI), $Bi_2Se_3$, exists even in the second BZ with unusually weak damping and narrow spectral linewidth, which likely relates to the spin-momentum locking feature of the TI surface states. This mode has significantly lower energy than that within the continuous medium excitation picture, indicating the existence of a bosonic coupling. Probably, the mode is dressed by a surface phonon due to strong electron-lattice interactions.

In the case of the Dirac states [8-10] at the surface of topological insulators (TIs), the unconventional spin textures protect the Dirac electrons against scattering from any non-magnetic impurity [11-13], resulting in a long lifetime for electrons in Dirac states. A relevant question is whether the bosonic collective modes from the electrons in these protected states, such as plasmons, have longer lifetimes than those from topologically trivial metallic states [14]. It has been proposed that the spin texture of Dirac electrons on a 3D TI surface may generate an undamped spin density wave, which can coherently couple to the charge density oscillations (*i.e.*, plasmon) and gives rise to the so-called spin-plasmon mode [15]. At the present time, the existing experimental results for plasmon modes on TI surfaces are rather controversial [16-19] and none of them provides direct evidence for the long lifetime characteristic of bosonic collective modes.

The single crystalline samples used here were Bi$_2$Se$_3$ [20]. The properties of the samples have been characterized by both transport and spectroscopic measurements [21]. The *in situ* cleaved (0001) surface is terminated with a Se layer and its *p*(1×1) structure is characterized by low energy electron diffraction (LEED), shown in Fig. 1(a). The band structure is confirmed by the mapping of both the linearly-dispersed topological Dirac surface states and parabolic bulk conducting band partly below the Fermi energy (Fig. 1(b)), using *in situ* angle-resolved photoemission spectroscopy (ARPES). The plasmons and other collective excitations, such as phonons, were measured with a newly developed high resolution electron energy loss spectroscopy (HREELS) system with the capability of two-dimensional (2D) energy and momentum mapping [22]. The HREELS measurements were performed along two high symmetry directions $\overline{\Gamma} - \overline{M}$ and $\overline{\Gamma} - \overline{K}$, as illustrated in the BZ shown in Fig. 1(a). Fig. 1(c) and 1(d) show the HREELS 2D mapping of Bi$_2$Se$_3$ (0001) along two high symmetry directions with several energy loss peaks labeled by α, OP, γ and ζ, respectively. Fig. 1(e) displays the energy distribution curves (EDCs) at $\overline{\Gamma}$, $\overline{K}$ and $\overline{M}$, respectively. The OP mode around 20 meV is almost dispersionless and assigned to an optical phonon of Bi$_2$Se$_3$, which corresponds to the $A_{1g}^2$ mode at Γ point in the Raman measurements [23] and is consistent with previous HREELS results [18]. The γ and ζ modes, around 55 meV and 72 meV without significant dispersions, are the conventional surface and bulk plasmon modes, respectively, originating from bulk conducting electrons [21]. These two modes, which have been observed but assigned to a single mode in Ref. [18], show strong damping and disappear at large *q*.

Distinct from the γ and ζ modes, the energy of α mode is strongly *q*-dependent. As shown in Fig. 1(f) and 1(g) for both energy-gain and loss sides, this α mode starts with zero energy at $\overline{\Gamma}$, disperses almost linearly along both $\overline{\Gamma} - \overline{K}$ and $\overline{\Gamma} - \overline{M}$ directions. Even entering into the second

BZ, this mode maintains the same linear dispersion without reflecting the lattice periodicity. This clearly indicates that the observed α mode is not an acoustic phonon, since the dispersion of an acoustic phonon must have the symmetry of the lattice. The observed dispersion of the α mode is the signature of plasmons.

To clarify the origin of the α mode, we performed HREELS measurements on Mn-doped $Bi_2Se_3$ (with 10% Mn doping). It is well-known that Mn substitutes for Bi upon doping in the crystal structure [24-26]. Mn doping changes the properties of both electronic bands and collective excitations of the system. Especially, such magnetic doping breaks the time reversal symmetry so that the Dirac surface band opens a gap. Heavy doping such as around 10% used here completely demolishes the Dirac surface band, as shown from the ARPES data in the inset of Fig. 2(a), which is consistent with previously reported results [27]. Changes of the plasmons and phonons can also be seen in our results. Fig. 2(a) shows the 2D HREELS energy-momentum mapping along the $\overline{\Gamma} - \overline{M}$ direction, while Fig. 2(b) displays the EDCs at the $\overline{\Gamma}$ point and $\overline{M}$ point. Compared with the HREELS results of $Bi_2Se_3$ (Fig. (1)), the conventional surface and bulk plasmon modes, γ and ζ, exhibit the same dispersion behaviors, but their energies increase from 55 meV to 94 meV and from 72 to 139 meV, respectively, owing to the increased bulk carrier density. This is evidenced by ARPES – the Fermi level is 0.15 eV above the conduction band minimum for $Bi_2Se_3$ (Fig. 1(b)), and 0.26 eV for Mn-doped $Bi_2Se_3$ (the inset of Fig. 2(a)). While changes of optical phonon (OP mode) caused by the Mn-doping are mild, only showing slight linewidth broadening (~0.7 meV) and energy shift (~1 meV).

However, the most intriguing change is the characteristics of the low-energy feature on the Mn-doped $Bi_2Se_3$ (denoted as AP for acoustic phonon). It is fundamentally different from the originally observed α plasmon mode for $Bi_2Se_3$, although they are similar in energy and dispersion in the low $q$ range. As clearly shown in Fig. 2(c), the energy of AP is symmetric with respect to the BZ, which is a signature of phonons. To summarize the difference, the dispersions of the α mode on Bi$_2$Se$_3$ and the AP mode on Mn-doped Bi$_2$Se$_3$ are plotted in Fig. 3(a). On the surface of Bi$_2$Se$_3$ where the topological surface states are present, the α mode appears, while the AP mode is suppressed. In contrast, on the surface of Mn-doped Bi$_2$Se$_3$ without the topological surface states, the α mode disappears and only the AP mode is present. Therefore, the α mode must originate from the topological surface states of Bi$_2$Se$_3$. In addition, from the measured dispersion curve of the AP mode [the pink curve in Fig3.(a)], we extracted the sound velocity $v_s = 1610 \pm 90$ m/s by a linear fit of the data points for small $q$. Notice that for acoustic phonons it is only possible to obtain transverse acoustic (TA) mode in HREELS measurements due to the restriction of the selection rules. Thus the AP mode is a TA mode with a transverse sound velocity of ~1610±90 m/s, which is very close to the reported value (1700 m/s) of Bi$_2$Se$_3$ [28,29].

The most striking behavior of the α mode is its slow attenuation in a large momentum range, which is in sharp contrast to the topologically trivial plasmon modes (*i.e.*, γ and ζ modes). As shown in Fig. 3(c) and 3(d), the ζ mode's intensity drops drastically by more than two orders of magnitude when dispersing into the corresponding EHPC at 0.016 Å$^{-1}$ and diminishes beyond 0.08 Å$^{-1}$, and its linewidth increases correspondingly. Other plasmon modes generated by topologically trivial surface states, *e.g.* the 2D plasmon on the surface of Si(111)-($\sqrt{3} \times \sqrt{3}$ )-Ag [30] or the acoustic plasmon on the surface of Be (0001) [31] show similar damping behavior. This is also true for the plasmon mode originated from Dirac electrons without topological protection, as is the case in graphene [7,32]. However, the α mode exhibits small decrease in intensity when dispersing across the EHPC (0.00005 Å$^{-1}$< $q$ < 0.208 Å$^{-1}$) and remains observable far beyond. Furthermore, its linewidth is almost momentum independent (Fig. 3(d)). In contrast to the AP mode which clearly shows multiple-scattering matrix effects [21], such characteristics of the α mode are intrinsic, independent of the incident electron energy. Moreover, the intensity of the AP mode at room temperature (300 K) is obviously higher than that at low temperature (35 K), while the intensity of the α mode is almost temperature-independent, strengthening the conclusion that the α mode observed in $Bi_2Se_3$ is not a simple phonon [21]. To understand why the α mode has an unusually long lifetime without fatal damping within the EHPC region, we should note that the Landau-type damping channels into the spin-unrestricted bulk band could be suppressed due to the spin-momentum locking restriction. Beyond the EHPC, the topological protection against any nonmagnetic impurity scattering also keeps the α mode from significant damping.

Another prominent characteristic of the α mode is its linear dispersion behavior in a large momentum range up to 2 Å$^{-1}$, independent of the periodicity of the lattice. However, the dispersion slope of the α mode (i.e., the velocity of the mode) is well over an order of magnitude smaller than that calculated from the continuous medium excitation picture [15,33,34]. This is also in sharp contrast to the case of the 2D plasmon mode from the Dirac electrons in graphene [7], where the dispersion is larger than theoretical calculations.

A possible interpretation of such anomalously small energy dispersion of the α mode is a coupling with other collective excitations, which results in a severe self-energy renormalization. Since its energy is close to that of the surface phonons, the α mode is likely to interact strongly with this excitation. This can be qualitatively understood by considering a plasmon dressed by an phonon mode due to strong electron-lattice interaction [35]. As a result, the plasmon frequency should be $\omega_q = \dfrac{\omega_{pl}}{\sqrt{\varepsilon_i(\omega,q)}}$, where $\omega_{pl}$ is the bare/undressed frequency, and $\varepsilon_i(\omega,q)$ the ionic effective dielectric function [36]. In a 2D electron liquid picture with short-range interactions [15,34], the bare plasmon was theoretically predicted to have acoustic dispersion (for $q < k_F$, the Fermi wave vector), $\omega_{pl} \approx v_F q$, where $v_F$ is the Fermi velocity. But these theories are based on random phase approximation so that they can only obtain the dispersions for small $q$, and neglect possible electron-lattice interactions. In Bi$_2$Se$_3$, a considerable number of studies [37-42] have shown both theoretically and experimentally that the surface Dirac states interact strongly with surface phonons, although the specific values of the electron phonon coupling constant and the interaction mechanism are still under debate [43]. With strong electron-lattice interaction, $\varepsilon_i(\omega, q) \propto \left( \Omega / \omega_i \right)^2 \gg 1$, where $\Omega$ represents the interaction strength and $\omega_i$ the phonon energy. Then the plasmon dispersion becomes $\omega_q \approx \left( \omega_i / \Omega \right) v_F q$, so it remains acoustic, but its slope (velocity) can be strongly reduced by electron-lattice interaction. Indeed, as shown in Fig. 1(f) and 1(g) as well as Fig. 3, the dispersion of the α mode is slightly anisotropic, e.g., the energy at $q = 1.2$ Å$^{-1}$ is 9.6 meV along the $\overline{\Gamma} - \overline{M}$ direction as compared to 8.5 meV along the $\overline{\Gamma} - \overline{K}$ direction. This indicates the effects of lattice symmetry which manifests the topological protection of the observed collective mode. However, whether or not the α mode couples with a spin density wave remains unclear. Our observations demonstrate both the surface Dirac states and their collective excitations are topologically protected on the 3D TI, though more quantitative studies are needed.

**Acknowledgments:**

We thank M. El-Batanouny, Z. Fang, Z. Wang, H. Weng, Y. Ran and C. Fang for helpful discussions. The work was supported by the National Key Research and Development Program of China (No. 2016YFA0302400 and No. 2016YFA0300600), the National Natural Science Foundation of China (No. 11304367, No. 11634016, and No. 11474334), and the Strategic Priority Research Program (B) of CAS (Grant No. XDB07010100). X.Z. was partially supported by the Youth Innovation Promotion Association of Chinese Academy of Sciences. J.Z. was partially supported by U.S. NSF under Grant No. DMR 1608865, and sabbatical program of the Institute of Physics, Chinese Academy of Sciences.

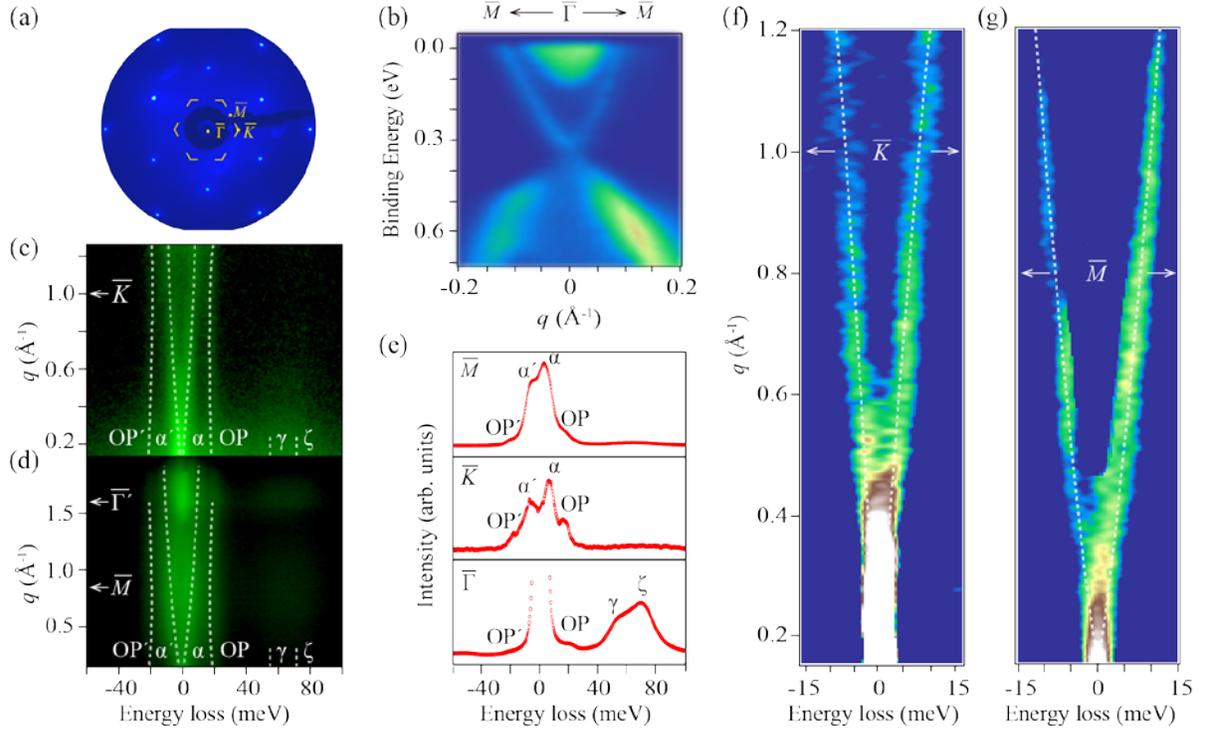

FIG. 1. Experimental results of Bi$_2$Se$_3$ (0001). (a) LEED pattern of Bi$_2$Se$_3$ (0001) surface with incident energy of 80 eV at room temperature. The red dashed line is the first BZ of Bi$_2$Se$_3$ (0001), and the red dots show the high-symmetry points. (b) The band structure along $\overline{M} - \overline{\Gamma} - \overline{M}$ direction measured by ARPES at 35 K, which shows the coexistence of bulk conducting band and the V-shaped surface Dirac state. (c) 2D energy-momentum mapping of HREELS along $\overline{\Gamma} - \overline{K}$ direction with incident energy of 60 eV and (d) along $\overline{\Gamma} - \overline{M}$ direction with incident energy of 110 eV. Four collective modes with energy loss features are labeled by α, OP, γ and ζ, respectively. The grey dash lines are provided to guide to eye. The corresponding negative energy loss features are anti-stocks peaks of α, OP modes respectively, which are labeled by α', OP'. (e) EDCs at $\overline{\Gamma}$, $\overline{K}$ and $\overline{M}$ with the corresponding energy loss peaks labeled. (f) and (g) The second derivative image of (c) and (d), respectively, with elastic background removed.

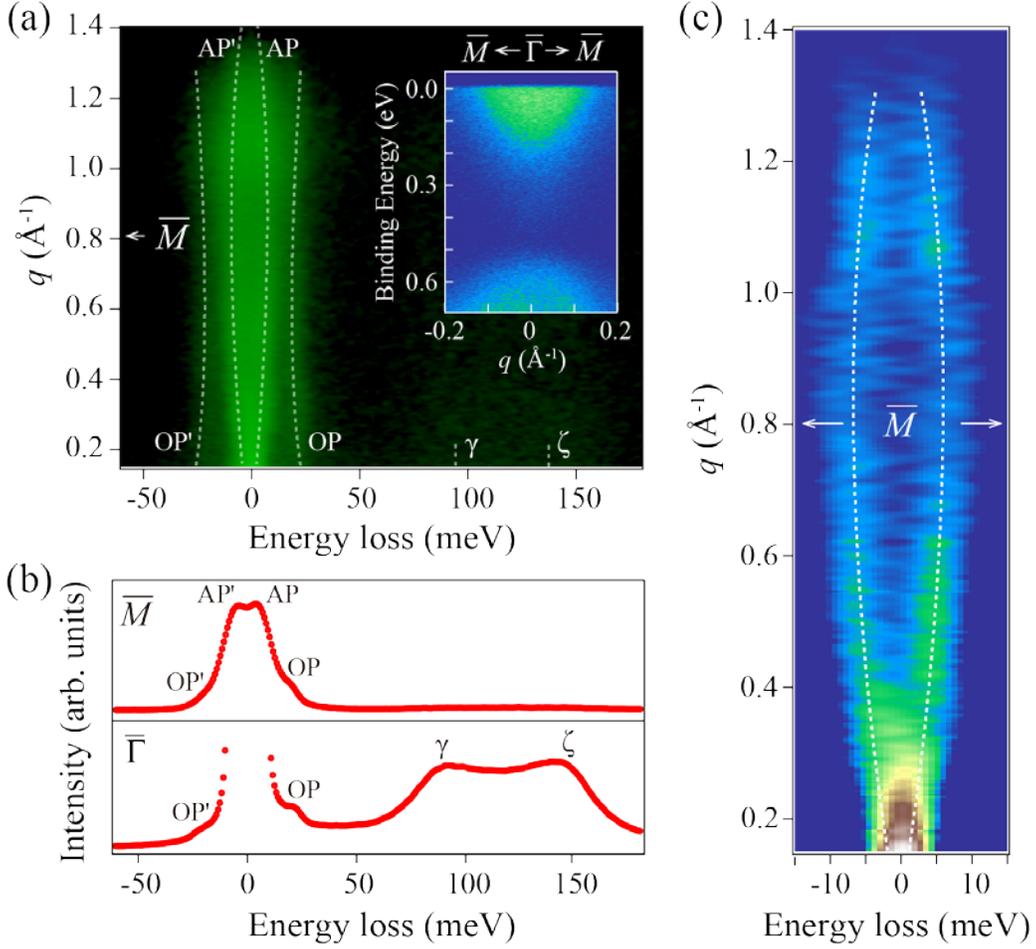

FIG. 2. Experimental results of Mn-doped $Bi_2Se_3$(0001). (a) 2D energy-momentum mapping of

HREELS measurements along $\overline{\Gamma} - \overline{M}$ direction with the incident energy of 110 eV at room temperature.

Four collective modes with energy loss features are labeled by AP, OP, $\gamma$ and $\zeta$ mode, respectively. The grey dash lines are provided to guide to eye. The corresponding negative energy loss features are anti-stocks peaks of AP, OP modes respectively, which are labeled by AP', OP'. The inset is the band structure along $\overline{M} - \overline{\Gamma} - \overline{M}$ direction measured by ARPES at 35 K, which shows that the V-shaped surface Dirac state disappears because of magnetic doping. (b) EDCs at $\overline{\Gamma}$ and $\overline{M}$ with corresponding energy loss peaks labeled. (c) The second derivative image of (a).

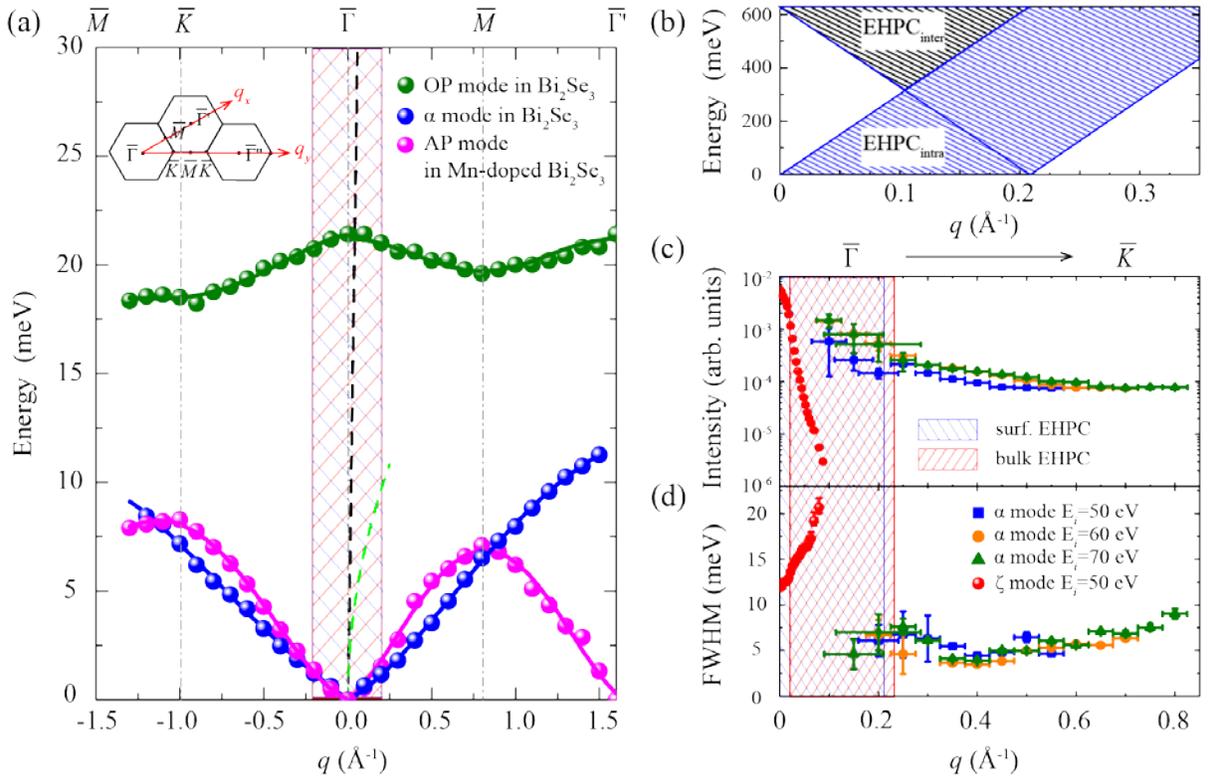

FIG. 3. Characteristics of the α plasmon mode. (a) The dispersions of α mode, OP mode of Bi$_2$Se$_3$ and AP mode of Mn-doped Bi$_2$Se$_3$. The dispersions of theoretically predicted Dirac plasmon and spin-plasmon are calculated by simply extrapolating the momentum range of the equations in Refs.[33] and [15], and plotted as black and green dashed lines, respectively. The shaded zones label the EHPC of massless Dirac electrons and normal bulk conducting electrons. (b) The enlarged EHPC of massless Dirac electrons in an extended energy range. (c) The normalized intensity and (d) FWHM of α mode of Bi$_2$Se$_3$ with different incident energies. The shaded zones illustrate the EHPC of Dirac surface electrons and normal bulk conducting electrons, respectively. The normalized intensity and FWHM of the bulk plasmon (ζ mode) of Bi$_2$Se$_3$ are also plotted for comparison.

Supplemental Materials for

# Anomalous Acoustic Plasmon Mode from Topologically Protected States

Xun Jia[1,2], Shuyuan Zhang[1,2], Raman Sankar[3,4], Fang-Cheng Chou[4], Weihua Wang[1], K. Kempa[5], E. W. Plummer[6], Jiandi Zhang[6], Xuetao Zhu[1,2*], Jiandong Guo[1,2,7*]

# 1    Materials and Methods

## 1.1    Crystal preparation

The single crystals of $Bi_2Se_3$ and $Mn_xBi_2Se_3$ used in present study were prepared by the vertical Bridgman growth method. The stoichiometric mixtures of high purity raw materials (5N) of bismuth, selenium and manganese for the $Bi_2Se_3$ and $Mn_xBi_2Se_3$ (x= 10 %) growth were grounded and then vacuum-sealed in quartz tubes of vacuum maintained at about $3 \times 10^{-2}$ torr. These initial mixtures in the vacuum-sealed tubes were heated at 923 K for 18 hours in a box furnace and then cooled slowly to room temperature. The pre-reacted powders were ground, vacuum-sealed in quartz tubes, and load into the vertical Bridgman furnace for the single crystal growth. A schematic growth set up is shown in the inset of Fig. S1. We have monitored and measured the temperature profile of the cooling process versus the relative position of sample, as shown in Fig. S1. The hot zone is maintained at 1173 K for 12 hours and a temperature gradient of about 0.7 K/mm is programmed near the solidification point of ~979 K (marked by the arrow in Fig. S1), and the quartz tubes are then moved slowly into the cooling zone with a translation rate of about 0.5 mm/hour.

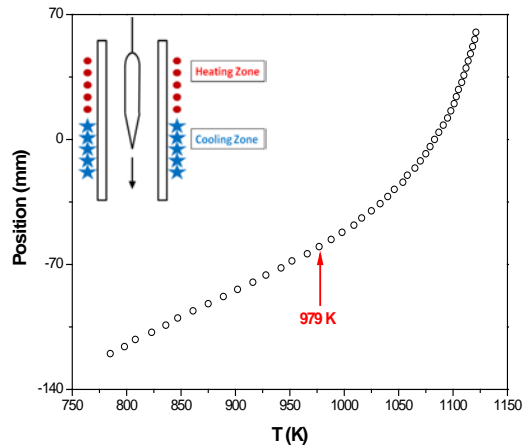

FIG. S1. The temperature profile of the cooling process versus the relative position of sample for the vertical Bridgman furnace shown in the inset.

## 1.2 Sample characterization

The properties of the samples were characterized by *ex situ* transport and spectroscopic measurements, as described in Ref. [S1]. The samples were cleaved *in situ* at room temperature to obtain fresh surface. The quality of the cleaved surface was checked *in situ* by LEED. Both $Bi_2Se_3$ and Mn-doped $Bi_2Se_3$ show very clear and sharp LEED pattern (shown in Fig. S2), indicating good quality of the cleaved surface. Meanwhile, the LEED pattern is used to easily determine the crystallographic direction for the HREELS measurements. The electronic band structure was measured by *in situ* ARPES as shown in the main text.

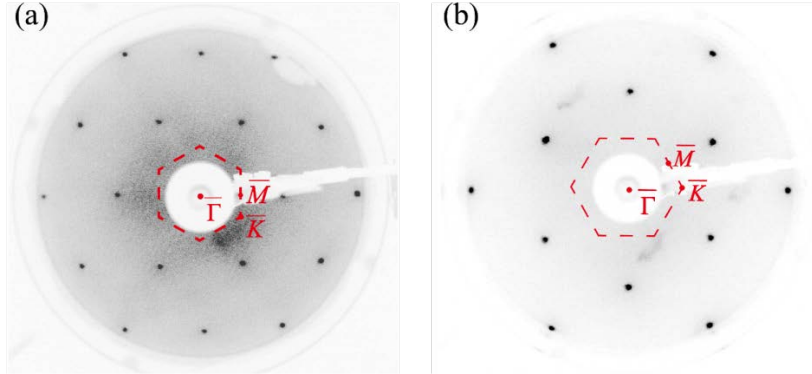

FIG. S2. (a) LEED pattern of Mn-doped $Bi_2Se_3$ (0001) surface with incident energy of 100 eV at room temperature. The red dashed line is the first BZ of Mn-doped $Bi_2Se_3$ (0001), and the three red dots are the high-symmetry points $\overline{\Gamma}$, $\overline{M}$, and $\overline{K}$, respectively. (b) LEED pattern of $Bi_2Se_3$ (0001) surface with incident energy of 80 eV at room temperature. The red dashed line is the first BZ of Mn-doped $Bi_2Se_3$ (0001), and the three red dots are the high-symmetry points $\overline{\Gamma}$, $\overline{M}$, and $\overline{K}$, respectively.

## 1.3 Brief introduction of 2D-HREELS system

As a surface sensitive technique, HREELS is an ideal candidate to explore the low-energy collective excitations at the surface of $Bi_2Se_3$. Compared with conventional HREELS, our recently developed 2D-HREELS can directly obtain a 2D energy-momentum mapping in a very large momentum scale without rotating sample, monochromator, or analyzer. A schematic illustration of the 2D-HREELS measurements is shown in Fig. S3.

The energy and momentum of the collective excitations (either plasmon or phonon) are obtained using the conservation of energy and momentum for incident and scattered electrons. As given by $\hbar q_{\parallel} = \hbar(k_i \sin\alpha_i - k_s \sin\alpha_s)$ (where $\alpha_i$ and $\alpha_s$ are the incident and scattering angles, respectively), the parallel momentum $q_{\parallel}$ depends on incident energy $E_i$, loss energy $E_{loss} = E_s - E_i$, $\alpha_i$ and $\alpha_s$ according to

$$q_{\parallel} = \frac{\sqrt{2mE_i}}{\hbar}\left(\sin\alpha_i - \sqrt{1 - \frac{E_{loss}}{E_i}}\sin\alpha_s\right) \approx \frac{\sqrt{2mE_i}}{\hbar}(\sin\alpha_i - \sin\alpha_s) \quad (S1)$$

(where $E_{loss} \ll E_i$), which means without changing other parameters we can directly obtain the information in a larger $q_{\parallel}$ range by increasing $E_i$. In this study, we obtain the information beyond the second Brillouin zone center $\overline{\Gamma'}$ along $\overline{\Gamma} - \overline{M} - \overline{\Gamma'}$ direction and beyond the $\overline{K}$ along $\overline{\Gamma} - \overline{K} - \overline{M}$ direction with the incident energy as large as 110 eV. All the HREELS measurements were performed within ~5 hours after the sample cleavage to make sure the data are collected from fresh surfaces.

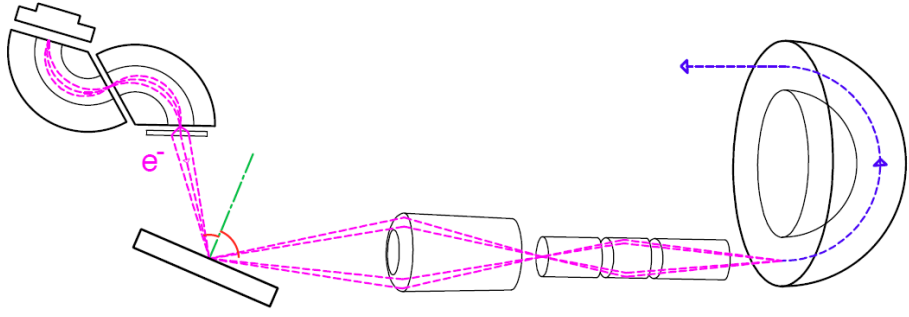

FIG. S3. Illustration of the 2D-HREELS measurement.

# 2 Identification of the trivial plasmon modes from bulk conducting band

The γ and ζ modes (in Fig. 1(c) and 1(d)), around 55 meV and 72 meV without significant dispersions, are the surface and bulk plasmons, respectively, originated from bulk conducting electrons.

Using the bulk carrier density $N_{3D} = 1.9 \times 10^{19} \text{cm}^{-3}$, effective mass $m^* \approx 0.19 m_e$ extracted from the measured band (Fig. 1(b)), and the dielectric constant $\varepsilon_r = 26$ [S2-S4], one can calculate the bulk plasmon energy $E_p = \hbar \omega_p = \hbar \sqrt{\dfrac{N_{3D} e^2}{\varepsilon_r \varepsilon_0 m^*}} \approx 73.2 \text{meV}$, and the energy of the corresponding surface plasmon $E_{sp} \approx \dfrac{E_p}{\sqrt{2}} \approx 51.8 \text{meV}$, consistent with the measured ζ and γ modes, respectively.

# 3 Fitting of the experimental results

The typical 2D-HREELS data is a mapping like Fig. 1(c) or Fig. 2(a). We can extract Energy Distribution Curves (EDCs) at different momentum and fit the curves by Guassian line shapes to obtain the energy, FWHM and intensity of the peaks. Fig. S4 shows a typical fitting case.

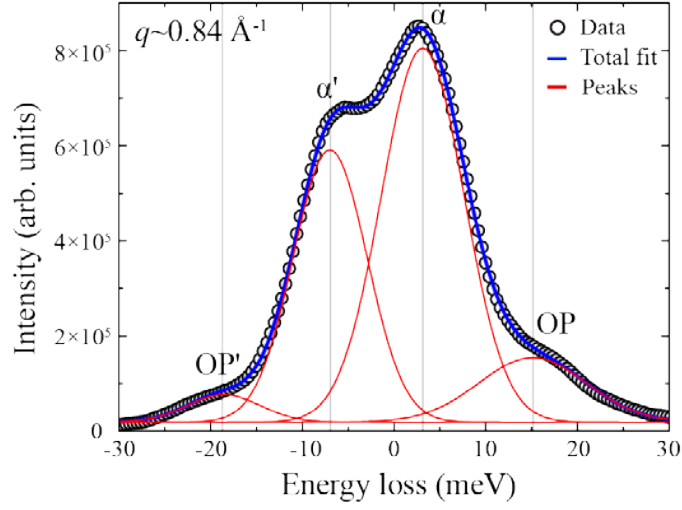

FIG. S4. Fitting the EDC at $q$=0.84 Å$^{-1}$ with energy loss region from -30 meV to 30 meV. Black circles are original data, red solid lines are Guassian peaks and blue solid lines are total fit.

# 4 The intensity ratio of stocks and anti-stocks peaks

According to Boltzmann statistics, the intensity ratio of the anti-stocks and stocks peaks can be expressed as:

$$\frac{I_{as}}{I_s} = e^{-\frac{E}{k_B T}}, \quad \text{(S2)}$$

where $I_{as}$ and $I_s$ are the intensity of anti-stocks and stocks peaks, respectively, $E$ is the energy of the detected boson (plasmon and phonon), $k_B$ is Boltzmann constant and $T$ is temperature.

Fig. S5a show the corresponding EDC stacking plots of Fig. 1(f). Fig. S5(b) shows the intensity ratio of anti-stocks peak and stocks peak of α mode, with the intensity data extracted from Fig. S5(a) and another two EDC stacking plots (no shown) with the incident energies of 70 eV and 110 eV. The experimental data fits well with the calculated curve.

(a)

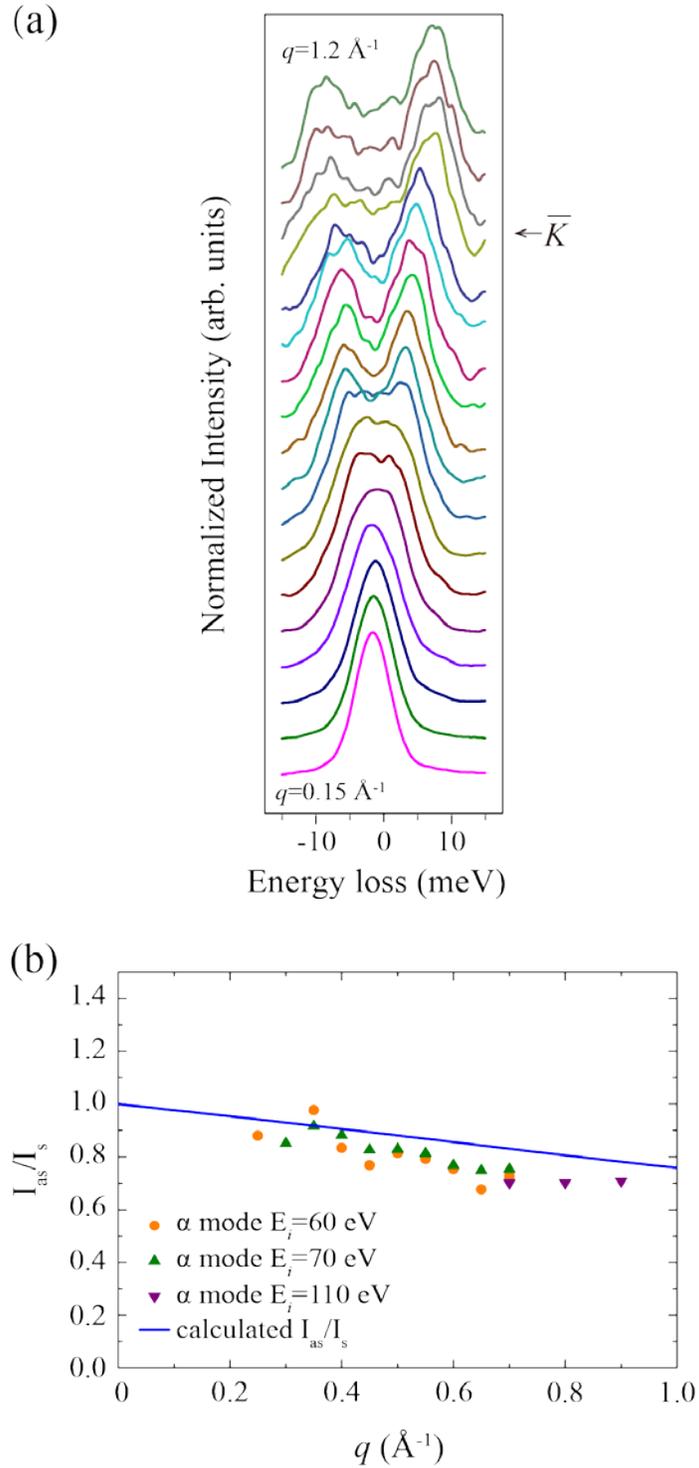

(b)

FIG. S5. (a) The corresponding EDC stacking plots of Fig. 1F. (b) Plot of the intensity ratio of anti-stocks peak and stocks peak of α mode as a function of momentum. Orange, olive, purple solid dots are the intensity ratio of α mode with the incident energies of 60 eV, 70 eV and 110 eV, respectively. And the blue line is the calculated curve using Eqn. (S2).

# 5 The intensity of AP mode in Mn-doped Bi₂Se₃

Compared to the normalized intensity of α mode in Bi₂Se₃ in Fig. 3(b), that of AP mode in Mn-doped Bi₂Se₃ is quite different. The normalized intensity with 60 eV is much larger than that of 80 eV (shown in Fig. S6), which is a signature of the multiple-scattering matrix effect of incident electrons. But the normalized intensity of α mode in Bi₂Se₃ is almost independent on incident energy (shown in Fig. 3(b)), which consolidates that AP mode in Mn-doped Bi₂Se₃ is a phonon and α mode in Bi₂Se₃ is indeed a plasmon associated with the inherent property of the TI surface.

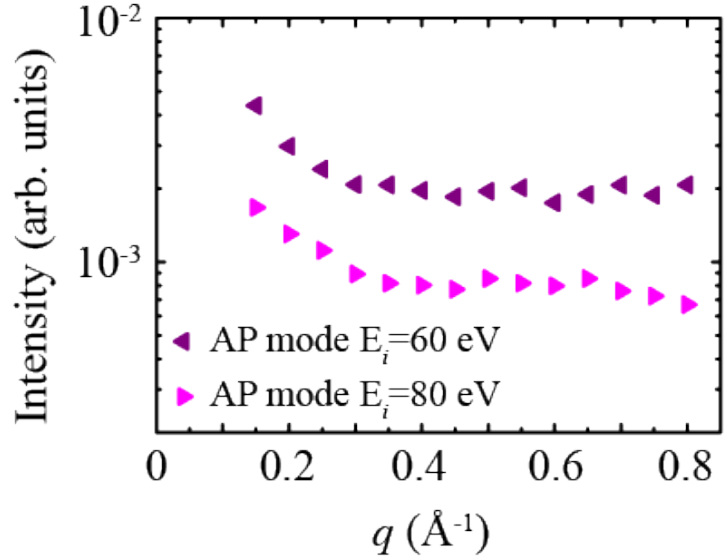

FIG. S6. The normalized intensity of AP mode in Mn-doped Bi₂Se₃ with different incident energies of 60 eV (wine solid dots) and 80 eV (pink solid dots) at 300 K.

# 6 More explanations of the $q$-dependence of α mode intensity

In the small $q$ range, the finite decrease of the α mode intensity is attributed to Landau-type damping, even though it is suppressed by spin-momentum locking restriction on the electron-hole pair excitation. The heavy overlapping between α mode and the elastic scattering peak due to limited energy resolution (~3 meV) prevents us from precisely resolve the intensity and linewidth of the mode, thus generating large error bars for the intensity as well as the linewidth in small $q$ range, i.e., the EHPC region (Fig. 3).

In order to investigate the intensity change (or damping of α mode) in a large momentum range, we performed the measurements with a larger incident energy (110 eV), with the results plotted in Fig. S7. We would like to mention that it takes quite a certain time to stabilize the electron beam when changing the beam energy. Therefore it is impossible to measure the necessary spectra from a single fresh sample surface using both the relative low energy (50, 60, and or 70 eV) with high resolution (as presented in the main manuscript) and the high energy (110 eV) to reach high $q$ range (presented in Fig. S7). The data were actually collected on different samples, while the overlapped range shows consistent results.

Beyond EHPC, the α mode intensity shows moderate decrease about one order of magnitude in a large momentum range of ~1.5 Å$^{-1}$, which is in sharp contrast to the topologically trivial plasmon modes. For example, the intensities of the bulk ζ mode in our study (Fig. 3(b)) and the 2D plasmon mode on the surface of Si(111)-($\sqrt{3}\times\sqrt{3}$)-Ag [S5] both decrease at least 3 orders of magnitude in a small momentum range of ~0.1 Å$^{-1}$.

The moderate decrease of α mode may come from two contributions. One is simply from the cross section effect of the electron scattering. The intensity of the dipole scattering naturally decreases with increasing $q$, which can be regarded as an extrinsic effect. The other contribution is an intrinsic effect since possible damping can still exist for the electrons in a surface state of 3D TIs. Unlike the 1D topological edge states in 2D TIs, where the back scatterings are fully prohibited, the surface states in 3D TIs is not completely protected since certain scattering channels other than back scattering are possible. Nevertheless, the weak decay without fatal damping observed in such a large momentum range reflects the topological protected nature of the mode.

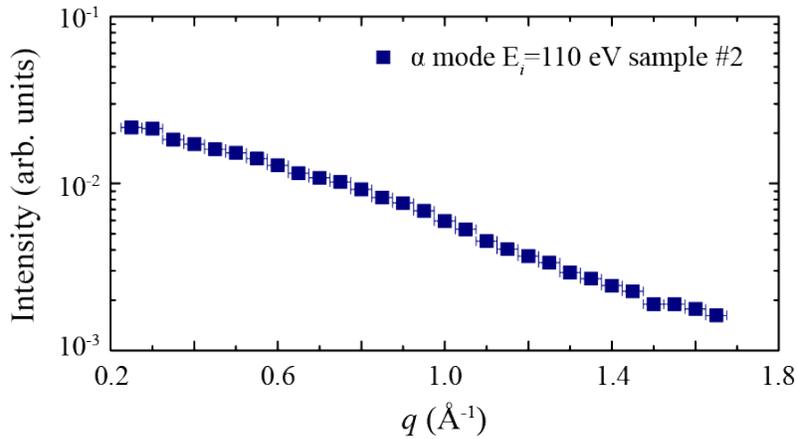

FIG. S7. The normalized intensity of α mode of Bi$_2$Se$_3$ with the incident energy of 110eV at 300 K.

# 7 Comparison of the temperature-dependent behavior of the mode intensity

In order to further investigate the different nature between the α mode of Bi$_2$Se$_3$ and the AP mode of Mn-doped Bi$_2$Se$_3$, we also performed HREELS measurements at low temperature (35 K). The dispersions of the two modes show no observable differences at 35 K compared to the results obtained at 300 K. However, the intensities of the two modes show different temperature-dependent behavior.

As shown in Fig.S8, the intensity of the AP mode at room temperature (300 K) is obviously higher than that at low temperature (35 K), which is an expected behavior of regular phonon characteristic. While the intensity of the α mode is almost temperature-independent, as expected from a plasmon mode, thus strengthening the conclusion that the α mode observed in $Bi_2Se_3$ is not a simple phonon.

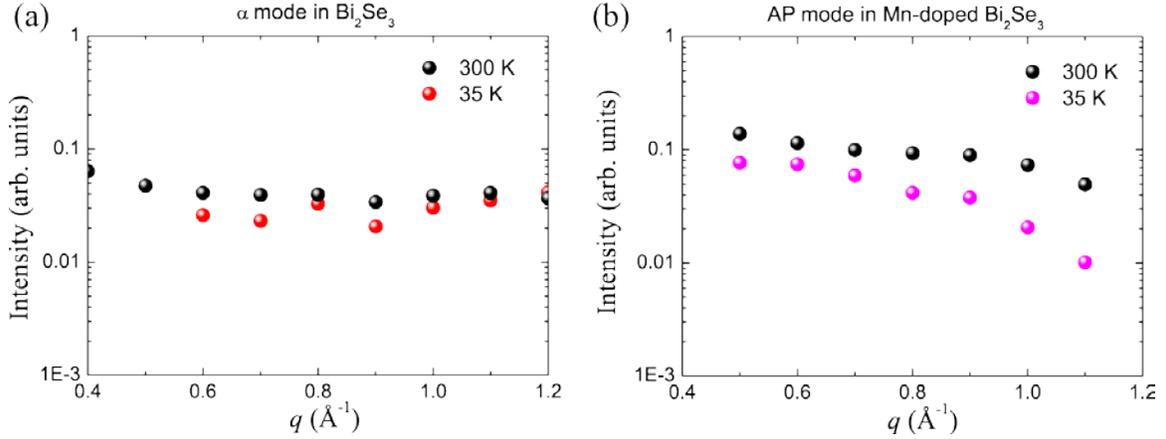

FIG. S8. Comparison of the temperature-dependent behavior between the α mode of $Bi_2Se_3$ (a) and the AP mode of Mn-doped $Bi_2Se_3$ (b).